\documentclass[a4paper,11pt]{article}
\usepackage{jcappub} 
\usepackage{lineno}
\usepackage{svg}

\usepackage{siunitx}
\usepackage{physics}
\usepackage{subcaption}
\usepackage{placeins}

\begin{document}

\title{A new Limit for Axion Dark Matter with SPACE}

\author[a]{M.~A.~Akgümüs,}
\author[a]{N.~Salama,}
\author[b]{J.~Egge,}
\author[a]{E.~Garutti,}
\author[a]{M.~Maroudas,}
\author[a]{L.~H.~Nguyen,}
\author[b]{D.~Leppla-Weber}

\affiliation[a]{University of Hamburg, 22761 Hamburg, Germany}
\affiliation[b]{Deutsches Elektronen-Synchrotron DESY, 22603 Hamburg, Germany}

\emailAdd{mehmet.akguemues@studium.uni-hamburg.de}
\emailAdd{nabil.salama@studium.uni-hamburg.de}
\abstract{
The axion, which has yet to be discovered, is a promising candidate for dark matter that emerges from Peccei-Quinn theory. This article presents the search for axion dark matter with the "Student Project for an Axion Cavity Experiment" (SPACE), which is also the first one in Germany. The hypothetical particle was looked for in the mass range from \SI{16.626}{\micro\electronvolt} to \SI{16.653}{\micro\electronvolt}, corresponding to a frequency range of 4.020~GHz to 4.027~GHz, using a resonant cavity in a peak magnetic field of 14~T. No significant signal was found, allowing us to exclude an axion-photon coupling $g_{a\gamma\gamma} = 14.6 \cdot 10^{-13}~\mathrm{GeV}^{-1}$ for the full mass range and $g_{a\gamma\gamma} = 2.8 \cdot 10^{-13}~\mathrm{GeV}^{-1}$ at peak sensitivity with a 95\% confidence level. This limit surpasses previous constraints by more than two orders of magnitude.
}

\maketitle
\flushbottom

\section{\label{sec:intro}Introduction}

Dark matter is thought to constitute most of the matter in the universe, but remains undetected due to its very weak interaction with ordinary matter. A hypothetical particle that could constitute all of dark matter is the axion ~\cite{Preskill:1982cy, Abbott:1982af, Dine:1982ah}, which emerges from Peccei-Quinn theory~\cite{Peccei:1977hh} and was first proposed by Wilczek~\cite{Wilczek:1977pj} and Weinberg~\cite{Weinberg:1977ma}. Its original purpose was to explain the non-observation of charge-parity (CP) violation in the strong sector, known as the strong CP-problem.
The axion mass $m_a$ is a priori unknown and could span many orders of magnitude. Astrophysical and cosmological observations put an upper bound of order $\mathcal{O}(\mathrm{eV})$~\cite{Marsh:2015xka}.
Under a static external magnetic field $\mathbf{B}_e$, axions convert into photons that can then be detected. The interaction between axions and photons is described by modified Maxwell's equations where Ampère's law has an additional term for the axion-induced current density~\cite{Sikivie:1983ip}:

\begin{equation}
\boldsymbol{\nabla} \times \boldsymbol{H} - \boldsymbol{\dot{D}} = \frac{g_{a\gamma\gamma}}{Z_0} \dot{a} \boldsymbol{B}_e,
\label{eq:ampere_law}
\end{equation}

with the magnetic field $\boldsymbol{H}$, the displacement field $\boldsymbol{D}$, the non-relativistic axion field $a$, the free space impedance $Z_0$ and the (unknown) axion-photon coupling constant $g_{a\gamma\gamma}$. The right-hand side of \autoref{eq:ampere_law} is often denoted as the axion-induced current density $\boldsymbol{J}_a = g_{a\gamma\gamma} \dot{a} \boldsymbol{B}_e / Z_0$. This current density can excite resonant electromagnetic modes of a cavity at a frequency corresponding to the axion mass $\nu_a \approx m_a c^2 / h$.
A uniform $\boldsymbol{J}_a$ that is aligned along the cylindrical cavity would predominantly excite the TM$_{010}$ mode.
The signal power extracted from the axion-induced resonant mode in a cavity with loaded quality factor $Q_L$, antenna coupling $\beta$, and resonance frequency $\nu$ is expected to be~\cite{Egge:2022gfp}

\begin{equation}
\begin{aligned}
\begin{split}
    P_s &= 1.7 \cdot 10^{-20}~\mathrm{W} \cdot \left( \frac{\beta}{1+\beta} \right)
    \left( \frac{V}{0.714~L} \right)
    \left( \frac{B_{\mathrm{rms}}}{12.66~\mathrm{T}} \right)^2
    \left( \frac{C_{010}}{0.682} \right) \\
    &\quad \times
    \left( \frac{g_{a\gamma\gamma}}{2.811 \cdot 10^{-13}~\mathrm{GeV}^{-1}} \right)^2 
    \left( \frac{16.64~\mathrm{\mu eV} }{m_a} \right)^2
    \left( \frac{\rho_a}{0.45~\mathrm{GeV/cm}^3} \right) \\
    &\quad \times
    \left( \frac{\nu_0}{4.023~\mathrm{GHz}} \right)
    \left( \frac{Q_L}{1.1 \cdot 10^4} \right)
    \left( \frac{1}{1 + \left( 2\delta\nu_a / \Delta\nu \right)^2} \right).
\end{split}
    \label{eq:axion_signal_power}
\end{aligned}
\end{equation}

Here $\delta\nu_a = \nu_0 - \nu_a$ is the detuning between the axion and cavity resonance frequencies, $\Delta\nu = \nu_0 / Q_L$ is the cavity mode linewidth, 
$C_{010}$ is the spatial overlap between the $\mathrm{TM}_{010}$ mode and $\boldsymbol{B}_e$ with a root mean squared value $B_\text{rms}$ over the cavity volume $V$. A local axion dark matter density $\rho_a = 0.45 \, \text{GeV} \, \text{cm}^{-3}$ \cite{local_DM_density} is assumed.

\section{Experimental Setup}

\begin{figure}[ht]
    \centering
    \includegraphics[height=0.12\textheight]{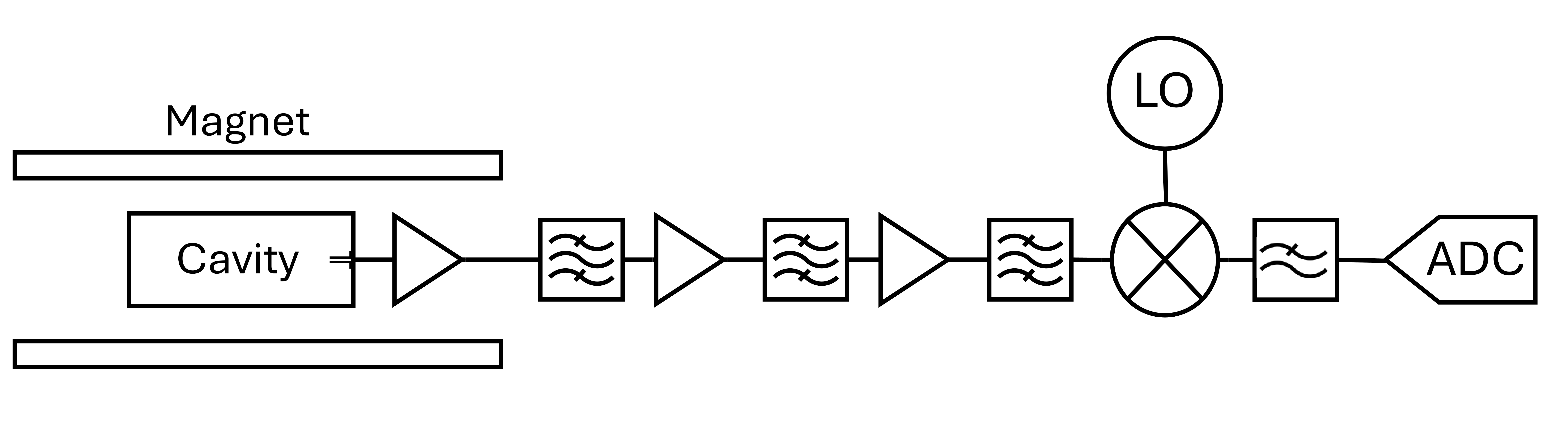}
    \caption[Cavity and receiver chain setup]{Cavity and receiver chain setup: Cylindrical resonant cavity in the magnet bore followed by a three-stage signal amplification receiver chain consisting of low noise amplifiers, bandpass filters and a mixing component for signal readout at the ADC.}

    \label{fig:receiver_chain}
\end{figure}

The experimental setup (\autoref{fig:receiver_chain}) consists of a resonant cavity placed in a solenoid superconducting electromagnet and a receiver chain coupled to the cavity via a dipole antenna.

\begin{figure}[ht]
    \centering
    \includegraphics[width=0.8\textwidth]{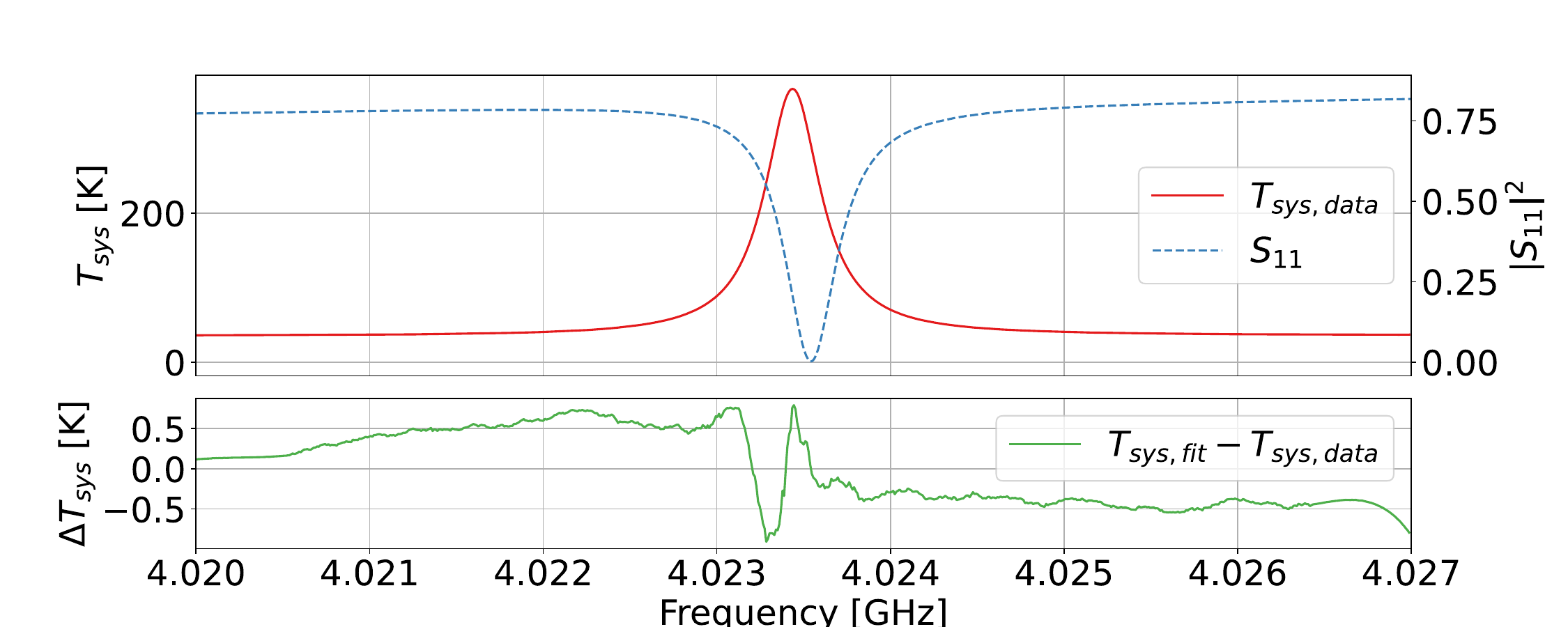}
    \caption[System noise temperature and reflection coefficient]{Reflection coefficient $\left|S_{11}\right|^2$ (dashed) and system noise temperature $T_\text{sys}$ (solid) of the setup as a function of frequency.
    The maximum signal power occurs at the minimum of $\left|S_{11}\right|^2$ and maximum of $T_\text{sys}$, respectively. It is slightly shifted between measurements because of changes in the physical environment. The residual $\Delta T_\text{sys}$ between the measured and fitted system temperature is shown in the lower plot.
    }
    \label{fig:T_sys_fit}
\end{figure}

The resonant cavity is a cylinder with an inner radius of $r = \SI{28.483}{\milli\meter}$ and a length of $l = \SI{280}{\milli\meter}$. It is made out of oxygen-free high conductivity copper with an inner surface roughness of \SIrange{0.2}{0.4}{\micro\meter}.
The resonance frequency $\nu_0$, quality factor $Q_L$ and coupling factor $\beta$ were determined by fitting a Lorentzian to a reflection measurement (\autoref{fig:T_sys_fit}) using a vector network analyzer\footnote{Rohde \& Schwarz ZVA 67}.
The measured resonance frequency of $\nu_0 = \SI{4.023}{\giga\hertz}$ for the $\mathrm{TM}_{010}$ mode is in accordance with the expected value
$
\hat{\nu}_0 = j_{0,1} c/2\pi r = \SI{4.023}{\giga\hertz},
$
where $j_{0,1}$ is the first root of the Bessel function $J_0(x)$.
The critically coupled dipole antenna has a coupling factor of $\beta = 1.07$. The loaded quality factor $Q_L = 11177$ is close to the expected value
$
\hat{Q}_L = \hat{Q}_0 / (1 + \beta) = 12005,
$
which is calculated using
\begin{equation}
    \hat{Q}_0 = \frac{j_{0,1}}{2(1 + r/l)} \sqrt{ \frac{\sigma_{\mathrm{Cu}}}{\pi \varepsilon_0 \hat{\nu}_0} },
    \label{eq:Q_0}
\end{equation}
where $\sigma_\text{Cu} = \SI{5.8e7}{S/m}$ is the conductivity of copper, and $\varepsilon_0$, $\mu_0$ are the vacuum permittivity and permeability respectively.
\\
A signal from the cavity is picked up by the critically coupled antenna and then amplified by a chain of low-noise-amplifiers (LNA). The first stage LNA is directly connected to the antenna and is oriented inside the bore such that the magnetic field does not affect the gain. Bandpass filters attenuate undesired signals outside the frequency range of interest between \SI{4.020}{\giga\hertz} and \SI{4.027}{\giga\hertz}. A mixer and a local oscillator running at \SI{3.905}{\giga\hertz} are used to down-convert to the intermediate frequency range (\SIrange{115}{122}{\mega\hertz}), which is within the ADC\footnote{Xilinx RFSoC4x2} measurement frequency range.
The receiver chain was calibrated using the Y-Factor method~\cite{y-factor-paper} by connecting a calibrated noise diode with an excess noise ratio of \SI{14.2}{\decibel} to the receiver chain. To avoid saturation of the amplifiers, an additional \SI{9.6}{\decibel} attenuator was placed between the receiver chain and the noise diode.
From the Y-factor calibration, an overall gain of $113.83 \pm 0.37~\si{\decibel}$ and an added noise temperature of $80 \pm 31~\si{\kelvin}$ was determined before the data taking. The overall system noise temperature $T_\text{sys}$ obtained from the power spectra during the data taking are averaged over the duration of the run, which is shown in \autoref{fig:T_sys_fit}. It ranges from \SI{36}{\kelvin} to a maximum of \SI{367}{\kelvin} at the resonance frequency.
\\\\
The extrema of the system noise temperature and reflection coefficient in \autoref{fig:T_sys_fit} differ slightly, which can be explained by the physical handling of the cavity between measurements resulting in minor distortions of the geometry. This does not affect the limit since the relevant variables are obtained from a fit of the power spectra taken during the experimental run.

\begin{figure}[ht]
  \centering
  \includegraphics[width=0.6\textwidth]{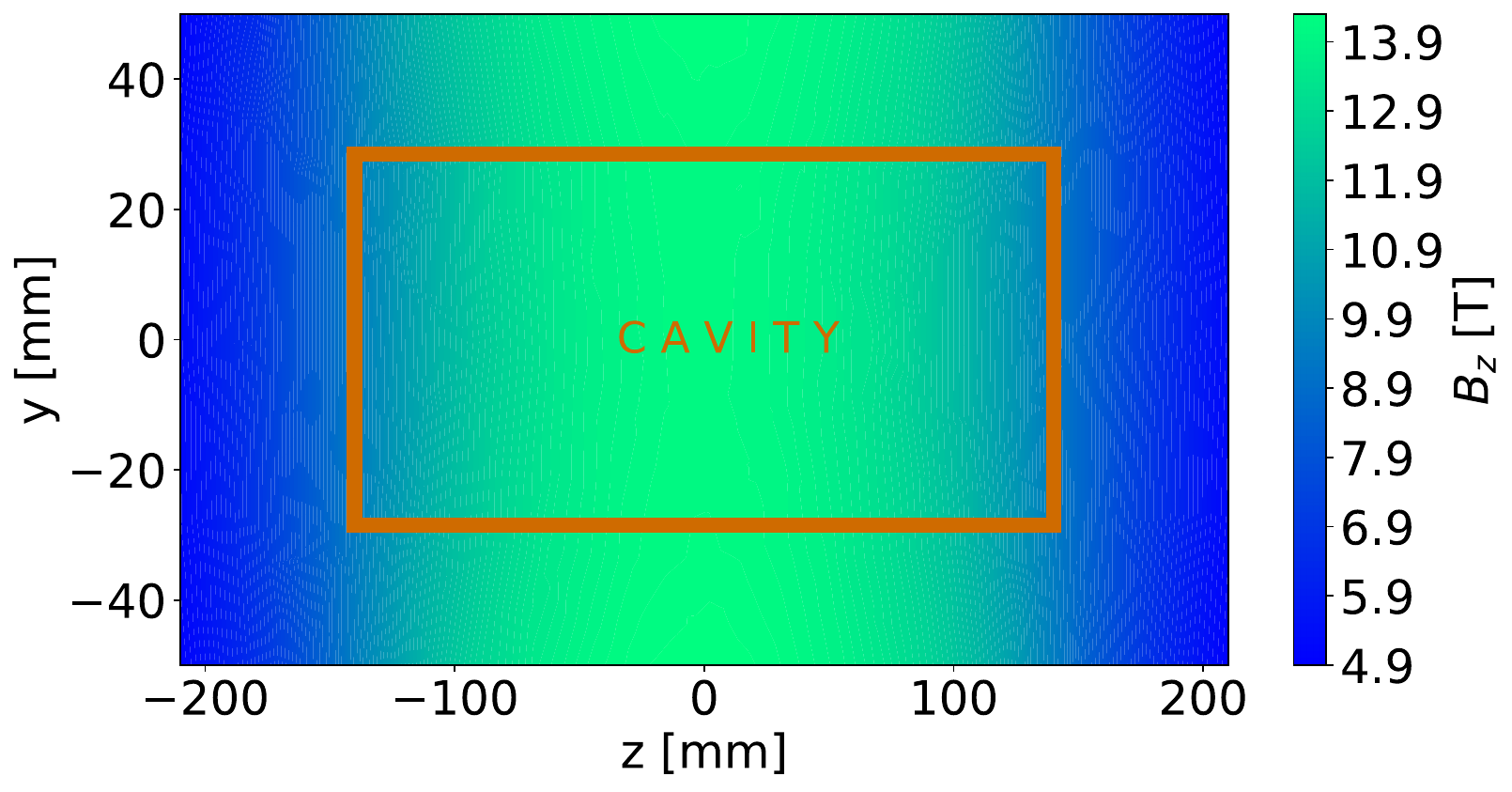}
  \caption[Magnetic field from simulation]{Magnetic field strength obtained from simulation inside and around the cavity (orange box) placed around $z=0$ such that the peak field strength of \SI{14}{\tesla} is achieved inside the cavity.}
  \label{fig:magnetic_field}
\end{figure}

The cavity is placed inside a superconducting electromagnet with a peak field strength of \SI{14}{\tesla} and a root mean square value $B_\text{rms} = \SI{12.66}{\tesla}$ over the cavity dimensions. The magnetic field was obtained from simulations of the electromagnet and is shown in \autoref{fig:magnetic_field}.

The form factor is defined as the overlap between the magnetic field $\mathbf{B}_e$ and the electric field of the resonant mode $\mathbf{E}_R$~\cite{Egge:2022gfp}:

\begin{equation}
  C_{010} = \frac{\left|\int_V dV\, \mathbf{E}_R \cdot \mathbf{B}_e \right|^2}{\int_V dV\, |\mathbf{B}_e|^2\, |\mathbf{E}_R|^2}.
  \label{eq:form_factor}
\end{equation}

For the electric field, the analytic expression for the TM\textsubscript{010} eigenmode was used. The integration is performed over the cavity volume $V$, resulting in $C_{010} = 0.682$. The uncertainty of the cavity position in the magnet has a negligible effect on $C_{010}$.

\section{Data Taking and Analysis}
The run was conducted from 19 April 2024 at 17:48 to 22 April 2024 at 17:34 at the location \ang{53;34;42.2}N \ang{9;53;09.6}E with the magnetic field pointing approximately at \ang{153} SE.
The run lasted 71 hours, during which 1.82 billion power spectra were obtained. The spectra were saved in batches of $2 \cdot 10^6$ spectra every $\sim$ 5 minutes. Considering the \SI{9.375}{\kilo\hertz} resolution bandwidth of the ADC, this results in an effective measurement time of \SI{54.4}{\hour}. Another run with no external magnetic field was also conducted, but not used for comparison in the analysis, since there were no significant signals found with the magnet on.

\begin{figure}[ht]
  \centering
  \includegraphics[width=0.8\textwidth]{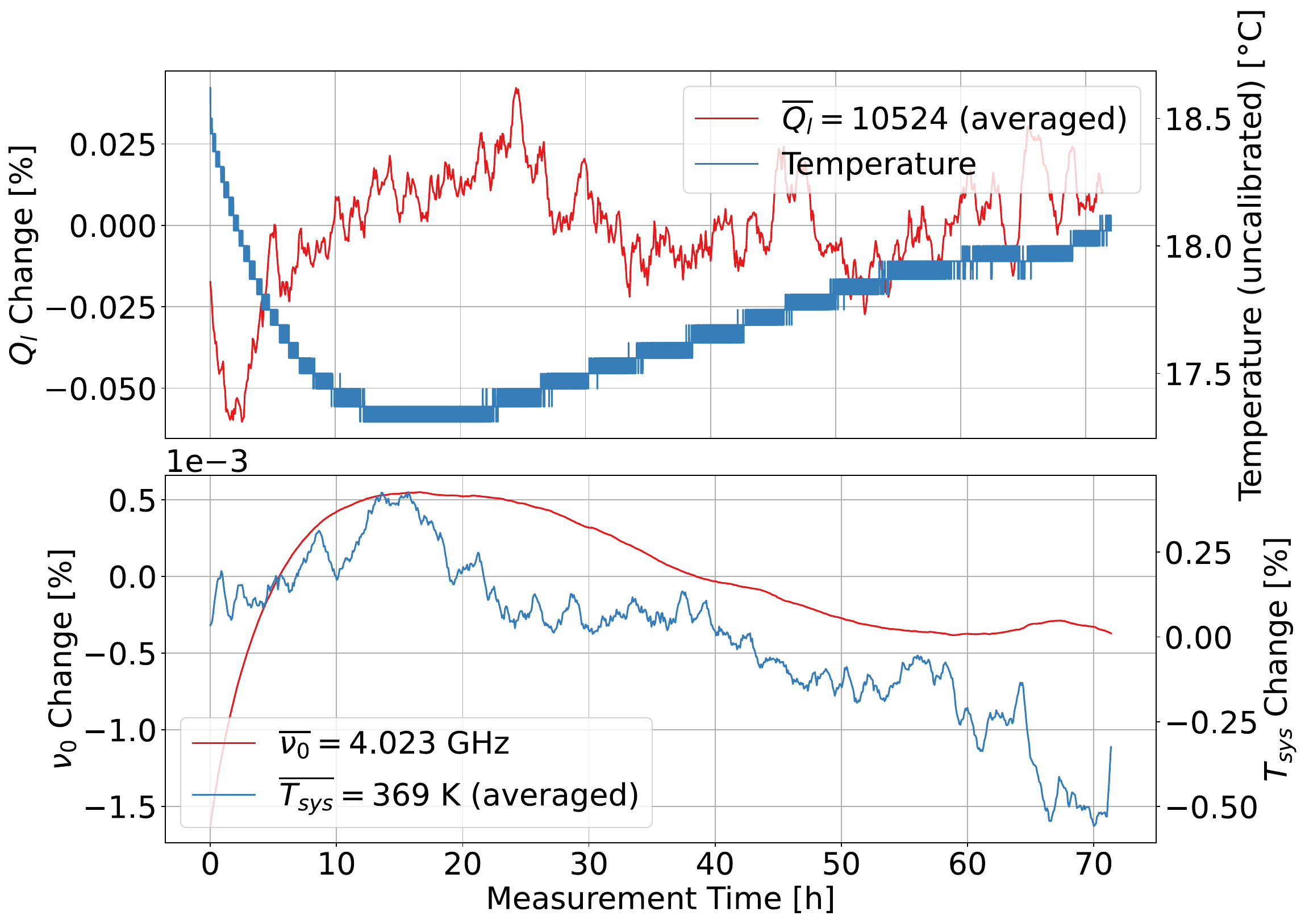}
  \caption[Temperature and cavity parameter evolution]{Change of physical temperature, loaded quality factor (running average), maximum system noise temperature  (running average), and resonance frequency during the run.}
  \label{fig:change_temp_freq}
\end{figure}

Environmental changes during the run led to drifts in the resonance frequency and quality factor of the cavity. In particular, operation of the magnet at \SI{4}{\kelvin} affects the temperature in the open bore.
By fitting a Lorentzian to each power spectrum, the resonance frequency and quality factor were extracted. The time evolution of these parameters, along with the peak system temperature and the physical temperature of the cavity, is shown in \autoref{fig:change_temp_freq}.
These changes were tracked and accounted for in the analysis. The correlation between physical temperature and cavity parameters is likely due to thermal contractions and improved conductivity of the cavity.

The procedure to optimally combine the saved spectra closely follows the HAYSTAC methodology~\cite{Brubaker:2017rna}. 
First, a 6th-degree Savitzky-Golay filter~\cite{savitzky-golay} with a \SI{178}{\kilo\hertz} window length is applied to each measured raw spectrum $P_{i,r}$ to obtain the filtered spectrum $P_{i,f}$. Any narrowband signal can then be revealed by subtracting and normalizing the raw spectra by the filtered spectra.
\begin{figure}[ht]
  \centering
  \includegraphics[width=0.65\textwidth]{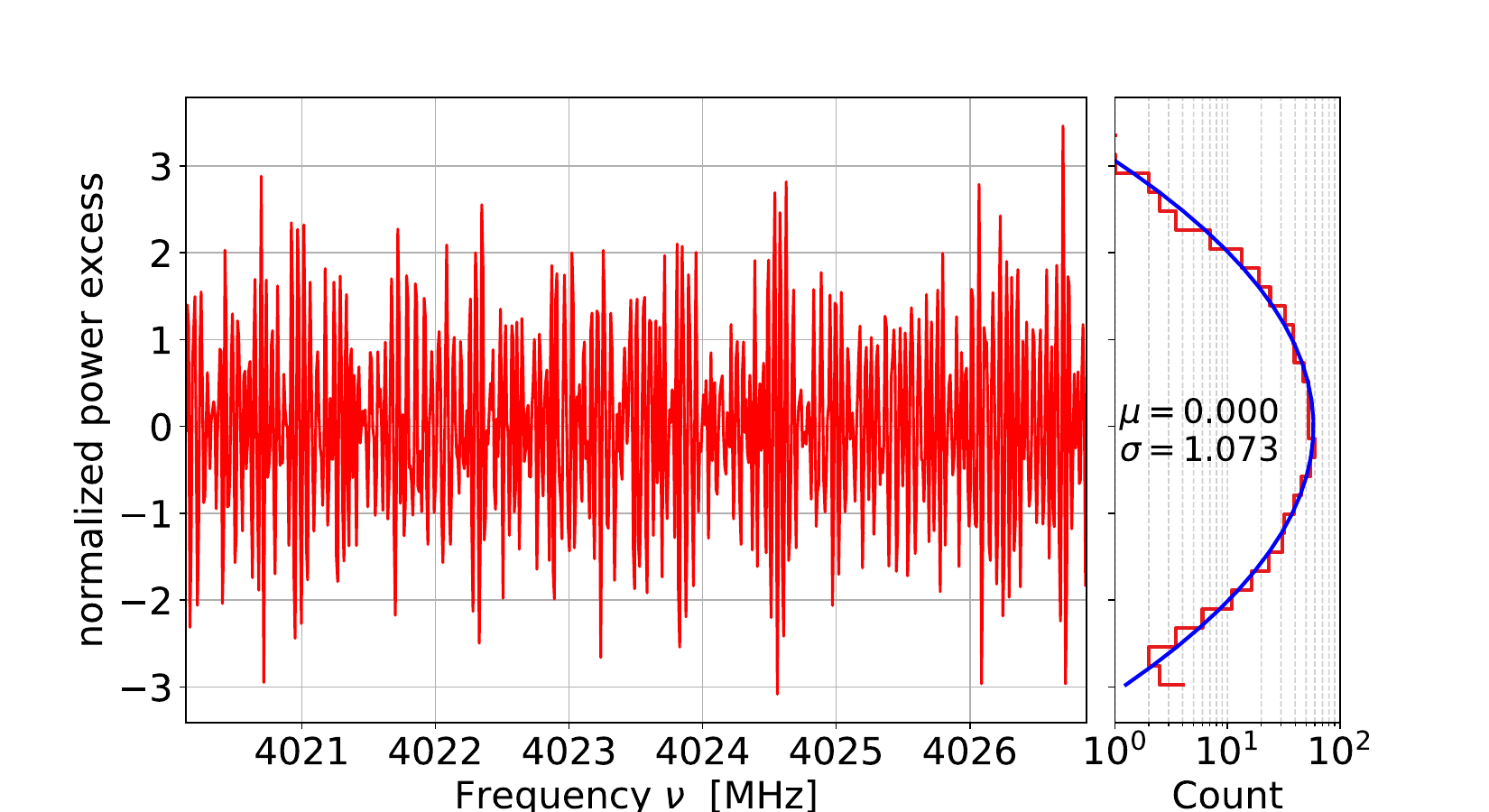}
  \caption[Grand spectrum]{Normalized power excess (grand spectrum) as a function of frequency. No significant excess is detected.}
  \label{fig:grand_spectrum}
\end{figure}
This yields the processed spectra $P_{i,p} = P_{i,r} / P_{i,f} - 1$, which are then combined using a weighted mean. The weights take into account the varying signal-to-noise ratio of a hypothetical axion signal due to the previously discussed changes in cavity parameters and system noise temperature~\cite{Brubaker:2017rna}. 
\\
To search for an axion signal, the resulting combined spectrum is cross-correlated with the expected signal shape, which is calculated using a standard boosted Maxwell-Boltzmann distribution~\cite{Diehl:2023fuk, OHare:2017yze} with a local dark matter velocity dispersion of $\sigma_v = \SI{154}{\kilo\meter\per\second}$~\cite{Bovy_2012_sigma_v}, and a laboratory relative velocity of $v_{\text{lab}} = \SI{242.1}{\kilo\meter\per\second}$~\cite{malhan2020_v_lab}.
As the resolution bandwidth is larger than the axion signal bandwidth of \SI{8.2}{\kilo\hertz}, the theoretical axion line shape is first convoluted by the frequency response of a single bin. With this, the different relative alignment between axion signal and each bin center frequency can systematically be treated. The worst-case scenario, in which the axion signal is split between two bins, is used.
\autoref{fig:grand_spectrum} shows the cross-correlated, normalized power excess, the so-called grand spectrum.
The largest excess found is $3.24 \sigma$, which corresponds to a global significance of $0.84\sigma$ or probability of $20.1\%$ of observing this excess due to random fluctuations. From the grand spectrum, a 95\% confidence level (CL) upper limit on the observed power is constructed for groups of 25 neighbouring bins using the approach described in~\cite{ADMX:2020hay}. Usage of the Savitzky-Golay filter attenuates a potential signal by $\eta_{\text{SG}} = 0.80$, which is determined by running the analysis on a total of 30000 simulated spectra with an injected axion signal. 
Taking this effect into account, the limit on signal power is converted into a limit on the axion-photon coupling $g_{a\gamma\gamma}$ using \autoref{eq:axion_signal_power}.

Systematic uncertainties are expected to be independent from each other and propagated onto the coupling via 
$
g_{a\gamma\gamma}^2 \propto \frac{T_{\text{sys}}}{Q_L} \cdot \frac{1+\beta}{\beta},
$
see \autoref{tab:propagated_errors}. The systematic uncertainties on the quality factor $Q_L$ and the antenna coupling $\beta$ stem from a difference between the calculation from the VNA measurement and the fit parameter from the measured power spectra during the run. 
For the system temperature $T_{\text{sys}}$, the uncertainty was obtained from the Y-factor calibration. The uncertainty on the form factor $C_{010}$ induced by the uncertainty of the cavity position inside the magnet is negligible. For the final limit, the most conservative value within its systematic uncertainty is adopted.

\begin{table}[htbp]
\centering
\begin{tabular}{lr|c}
\hline
\textbf{error effect} & \textbf{relative error} & \textbf{propagated uncertainty} \\
\hline
quality factor $Q_L$ & 5.9\% & 3.0\% \\
antenna coupling $\beta$ & 5.9\% & 1.4\% \\
system temperature $T_{\text{sys}}$ & 8.6\% & 4.3\% \\
\hline
Total error $g_{a\gamma\gamma} \propto \sqrt{\frac{1+\beta}{\beta} \cdot \frac{T_{\text{sys}}}{Q_L}}$ & & 5.4\% \\
\hline
\end{tabular}
\caption{Propagated uncertainties on the axion-photon coupling.\label{tab:propagated_errors}}
\end{table}

\begin{figure}[ht]
  \centering
  \includegraphics[height=0.3\textheight]{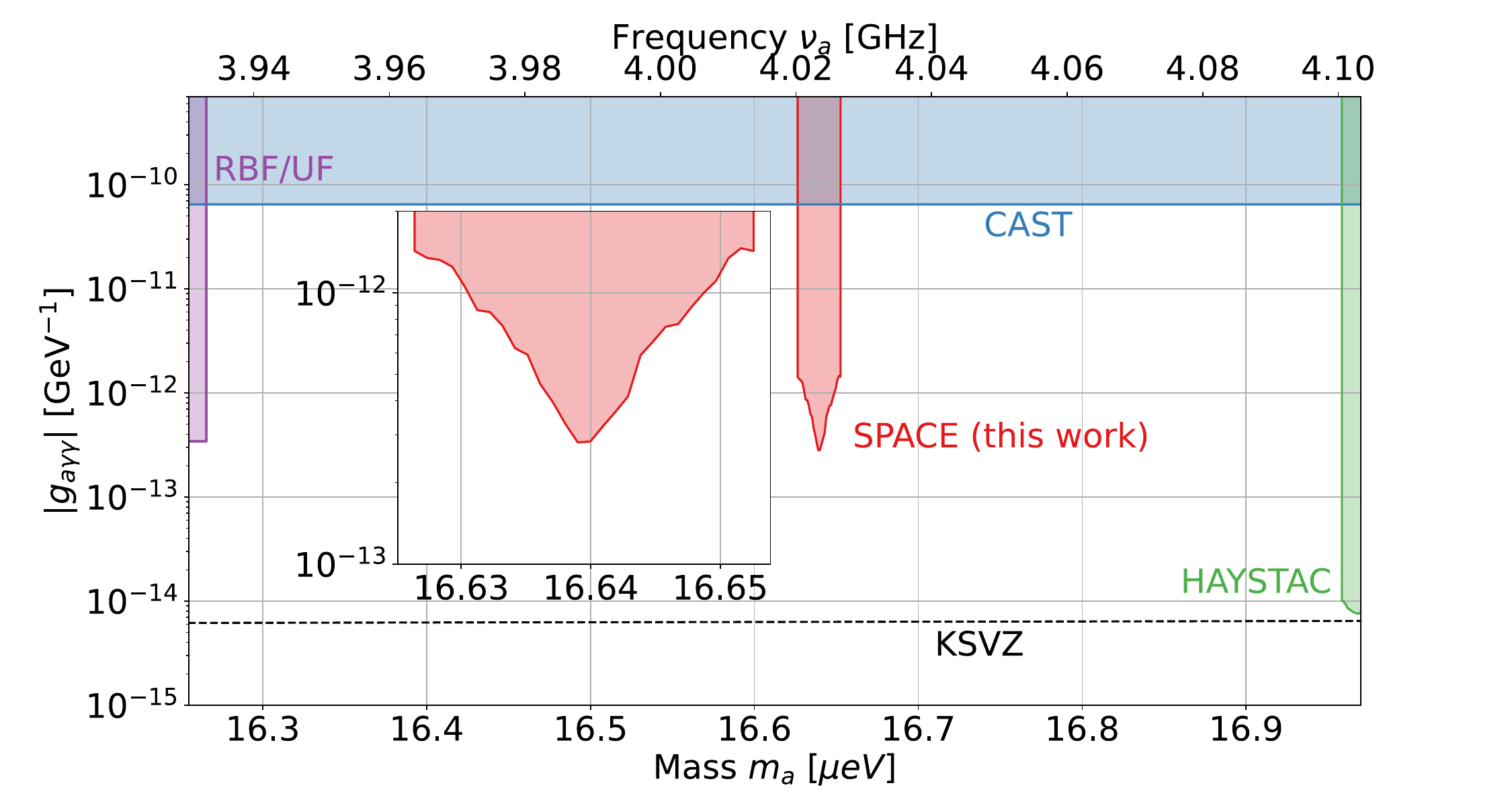} 
  \caption[Axion exclusion limit]{Axion exclusion limit obtained in this work. Exclusion for masses between $16.626\,\si{\micro\eV}$ to $16.653\,\si{\micro\eV}$ with coupling reaching $g_{a\gamma\gamma} = 2.8 \cdot 10^{-13}\,\si{\giga\eV}^{-1}$. Exclusion limits of other experiments are also shown~\cite{CAST:2024eil, HAYSTAC:2020kwv, Wuensch:1989sa}. All limits are plotted assuming a dark matter density $\rho_a = \SI{0.45}{\giga\eV\per\cubic\centi\meter}$.}
  \label{fig:grand_limit}
\end{figure}

\section{Conclusion}

The 95\% CL exclusion limit on the axion-photon coupling $g_{a\gamma\gamma}$ is shown in \autoref{fig:grand_limit}. 
Axion dark matter with a local density of $\rho_a = \SI{0.45}{\giga\eV\per\cubic\centi\meter}$ is excluded for coupling strengths larger than 
$g_{a\gamma\gamma} = 14.6 \cdot 10^{-13}\,\si{\giga\eV}^{-1}$ 
over the mass range from $16.626\,\si{\micro\eV}$ to $16.653\,\si{\micro\eV}$. 
At peak sensitivity, values of 
$g_{a\gamma\gamma} > 2.8 \cdot 10^{-13}\,\si{\giga\eV}^{-1}$ 
are excluded.

This result surpasses previous constraints in this mass range, including the CAST limit~\cite{CAST:2024eil}, by more than two orders of magnitude and is within a factor of 44 of the KSVZ benchmark sensitivity~\cite{KSVZ-1, KSVZ-2, GrillidiCortona:2015jxo}.

\begin{acknowledgments}
We would like to thank the HUB for Crossdisciplinary Learning (HCL) of the University of Hamburg for selecting us and this project for their funding program.
Additionally, we appreciate the support and resources provided by the Department for Experimental Physics of the University of Hamburg and their workshop for the construction of the cavity and usage of the magnet.
We thank the MADMAX team in Hamburg for their support and providing vital hardware.
We also thank Stefan Knirck and the BREAD collaboration for their advice and the firmware for the DAQ.

This project is partially funded by the Federal Ministry of Education and Research (BMBF) and the Free and Hanseatic City of Hamburg under the Excellence Strategy of the Federal Government and the Länder.
We acknowledge support by the Helmholtz Association (Germany).
Part of this project is funded by the Deutsche Forschungsgemeinschaft (DFG, German Research Foundation) under Germany’s Excellence Strategy – EXC 2121 “Quantum Universe” – 390833306, and through the DFG funds for major instrumentation grant DFG INST 152/8241.
\end{acknowledgments}


\bibliographystyle{JHEP}
\bibliography{references.bib}

\providecommand{\noopsort}[1]{}\providecommand{\singleletter}[1]{#1}%

\providecommand{\href}[2]{#2}\begingroup\raggedright\begin{thebibliography}{10}

\bibitem{Preskill:1982cy}
J.~Preskill, M.B.~Wise and F.~Wilczek, \emph{{Cosmology of the Invisible Axion}}, \href{https://doi.org/10.1016/0370-2693(83)90637-8}{\emph{Phys. Lett. B} {\bfseries 120} (1983) 127}.

\bibitem{Abbott:1982af}
L.F.~Abbott and P.~Sikivie, \emph{{A Cosmological Bound on the Invisible Axion}}, \href{https://doi.org/10.1016/0370-2693(83)90638-X}{\emph{Phys. Lett. B} {\bfseries 120} (1983) 133}.

\bibitem{Dine:1982ah}
M.~Dine and W.~Fischler, \emph{{The Not So Harmless Axion}}, \href{https://doi.org/10.1016/0370-2693(83)90639-1}{\emph{Phys. Lett. B} {\bfseries 120} (1983) 137}.

\bibitem{Peccei:1977hh}
R.D.~Peccei and H.R.~Quinn, \emph{{CP Conservation in the Presence of Instantons}}, \href{https://doi.org/10.1103/PhysRevLett.38.1440}{\emph{Phys. Rev. Lett.} {\bfseries 38} (1977) 1440}.

\bibitem{Wilczek:1977pj}
F.~Wilczek, \emph{{Problem of Strong $P$ and $T$ Invariance in the Presence of Instantons}}, \href{https://doi.org/10.1103/PhysRevLett.40.279}{\emph{Phys. Rev. Lett.} {\bfseries 40} (1978) 279}.

\bibitem{Weinberg:1977ma}
S.~Weinberg, \emph{{A New Light Boson?}}, \href{https://doi.org/10.1103/PhysRevLett.40.223}{\emph{Phys. Rev. Lett.} {\bfseries 40} (1978) 223}.

\bibitem{Marsh:2015xka}
D.J.E.~Marsh, \emph{{Axion Cosmology}}, \href{https://doi.org/10.1016/j.physrep.2016.06.005}{\emph{Phys. Rept.} {\bfseries 643} (2016) 1} [\href{https://arxiv.org/abs/1510.07633}{{\ttfamily 1510.07633}}].

\bibitem{Sikivie:1983ip}
P.~Sikivie, \emph{{Experimental Tests of the Invisible Axion}}, \href{https://doi.org/10.1103/PhysRevLett.51.1415}{\emph{Phys. Rev. Lett.} {\bfseries 51} (1983) 1415}.

\bibitem{Egge:2022gfp}
J.~Egge, \emph{{Axion haloscope signal power from reciprocity}}, \href{https://doi.org/10.1088/1475-7516/2023/04/064}{\emph{JCAP} {\bfseries 04} (2023) 064} [\href{https://arxiv.org/abs/2211.11503}{{\ttfamily 2211.11503}}].

\bibitem{local_DM_density}
J.I.~Read, \emph{{The Local Dark Matter Density}}, \href{https://doi.org/10.1088/0954-3899/41/6/063101}{\emph{J. Phys. G} {\bfseries 41} (2014) 063101} [\href{https://arxiv.org/abs/1404.1938}{{\ttfamily 1404.1938}}].

\bibitem{y-factor-paper}
D.~Pozar, \emph{Microwave engineering},  2011.

\bibitem{Brubaker:2017rna}
B.M.~Brubaker, L.~Zhong, S.K.~Lamoreaux, K.W.~Lehnert and K.A.~van Bibber, \emph{{HAYSTAC axion search analysis procedure}}, \href{https://doi.org/10.1103/PhysRevD.96.123008}{\emph{Phys. Rev. D} {\bfseries 96} (2017) 123008} [\href{https://arxiv.org/abs/1706.08388}{{\ttfamily 1706.08388}}].

\bibitem{savitzky-golay}
A.~Savitzky and M.J.E.~Golay, \emph{Smoothing and differentiation of data by simplified least squares procedures.}, \href{https://doi.org/10.1021/ac60214a047}{\emph{Analytical Chemistry} {\bfseries 36} (1964) 1627} [\href{https://arxiv.org/abs/https://doi.org/10.1021/ac60214a047}{{\ttfamily https://doi.org/10.1021/ac60214a047}}].

\bibitem{Diehl:2023fuk}
J.~Diehl, J.~Knollm\"uller and O.~Schulz, \emph{{Bias-free estimation of signals on top of unknown backgrounds}}, \href{https://doi.org/10.1016/j.nima.2024.169259}{\emph{Nucl. Instrum. Meth. A} {\bfseries 1063} (2024) 169259} [\href{https://arxiv.org/abs/2306.17667}{{\ttfamily 2306.17667}}].

\bibitem{OHare:2017yze}
C.A.J.~O'Hare and A.M.~Green, \emph{{Axion astronomy with microwave cavity experiments}}, \href{https://doi.org/10.1103/PhysRevD.95.063017}{\emph{Phys. Rev. D} {\bfseries 95} (2017) 063017} [\href{https://arxiv.org/abs/1701.03118}{{\ttfamily 1701.03118}}].

\bibitem{Bovy_2012_sigma_v}
J.~Bovy, C.A.~Prieto, T.C.~Beers, D.~Bizyaev, L.N.~da~Costa, K.~Cunha et~al., \emph{{The Milky Way's circular-velocity curve between 4 and 14kpc from apogee data}}, \href{https://doi.org/10.1088/0004-637X/759/2/131}{\emph{Astrophys. J.} {\bfseries 759} (2012) 131} [\href{https://arxiv.org/abs/1209.0759}{{\ttfamily 1209.0759}}].

\bibitem{malhan2020_v_lab}
K.~Malhan, R.A.~Ibata and N.F.~Martin, \emph{{Measuring the Sun's velocity using Gaia EDR3 observations of Stellar Streams}},  \href{https://arxiv.org/abs/2012.05271}{{\ttfamily 2012.05271}}.

\bibitem{ADMX:2020hay}
{\scshape ADMX} collaboration, \emph{{Axion dark matter experiment: Run 1B analysis details}}, \href{https://doi.org/10.1103/PhysRevD.103.032002}{\emph{Phys. Rev. D} {\bfseries 103} (2021) 032002} [\href{https://arxiv.org/abs/2010.06183}{{\ttfamily 2010.06183}}].

\bibitem{CAST:2024eil}
{\scshape CAST} collaboration, \emph{{New Upper Limit on the Axion-Photon Coupling with an Extended CAST Run with a Xe-Based Micromegas Detector}}, \href{https://doi.org/10.1103/PhysRevLett.133.221005}{\emph{Phys. Rev. Lett.} {\bfseries 133} (2024) 221005} [\href{https://arxiv.org/abs/2406.16840}{{\ttfamily 2406.16840}}].

\bibitem{HAYSTAC:2020kwv}
{\scshape HAYSTAC} collaboration, \emph{{A quantum-enhanced search for dark matter axions}}, \href{https://doi.org/10.1038/s41586-021-03226-7}{\emph{Nature} {\bfseries 590} (2021) 238} [\href{https://arxiv.org/abs/2008.01853}{{\ttfamily 2008.01853}}].

\bibitem{Wuensch:1989sa}
W.~Wuensch, S.~De~Panfilis-Wuensch, Y.K.~Semertzidis, J.T.~Rogers, A.C.~Melissinos, H.J.~Halama et~al., \emph{{Results of a Laboratory Search for Cosmic Axions and Other Weakly Coupled Light Particles}}, \href{https://doi.org/10.1103/PhysRevD.40.3153}{\emph{Phys. Rev. D} {\bfseries 40} (1989) 3153}.

\bibitem{KSVZ-1}
J.E.~Kim, \emph{{Weak Interaction Singlet and Strong CP Invariance}}, \href{https://doi.org/10.1103/PhysRevLett.43.103}{\emph{Phys. Rev. Lett.} {\bfseries 43} (1979) 103}.

\bibitem{KSVZ-2}
M.A.~Shifman, A.I.~Vainshtein and V.I.~Zakharov, \emph{{Can Confinement Ensure Natural CP Invariance of Strong Interactions?}}, \href{https://doi.org/10.1016/0550-3213(80)90209-6}{\emph{Nucl. Phys. B} {\bfseries 166} (1980) 493}.

\bibitem{GrillidiCortona:2015jxo}
G.~Grilli~di Cortona, E.~Hardy, J.~Pardo~Vega and G.~Villadoro, \emph{{The QCD axion, precisely}}, \href{https://doi.org/10.1007/JHEP01(2016)034}{\emph{JHEP} {\bfseries 01} (2016) 034} [\href{https://arxiv.org/abs/1511.02867}{{\ttfamily 1511.02867}}].

\end{thebibliography}\endgroup

\end{document}